\documentclass[10pt,conference]{IEEEtran}
\IEEEoverridecommandlockouts

\usepackage[numbers]{natbib}
\usepackage{amsmath,amssymb,amsfonts}
\usepackage{algorithmic}
\usepackage{booktabs}
\usepackage{graphicx}
\usepackage{mathrsfs}
\usepackage{multirow}
\usepackage{textcomp}
\usepackage{xcolor}
\usepackage{tcolorbox}
\usepackage{hyperref}
\usepackage{pifont}
\usepackage{enumitem}
\usepackage[framemethod=tikz]{mdframed}
\mdfdefinestyle{customstyle}{
  linecolor=gray!100,
  linewidth=3pt,
  innerleftmargin=3pt,
  topline=false,
  rightline=false,
  bottomline=false,
  leftline=true,
  innerrightmargin=3pt,
  innertopmargin=3pt,
  innerbottommargin=3pt,
  backgroundcolor=gray!15
}
\newenvironment{custommdframed}
  {\begin{mdframed}[style=customstyle]}
  {\end{mdframed}}

\def\BibTeX{{\rm B\kern-.05em{\sc i\kern-.025em b}\kern-.08em
    T\kern-.1667em\lower.7ex\hbox{E}\kern-.125emX}}
\begin{document}

\newcommand{\MethodName}{\textsc{FLAME}}
\newcommand{\DatasetName}{\textsc{PADefects}}

\newcommand{\fang}[1]{{\color{blue}#1}}

\newcommand{\jing}[1]{{\color{purple}#1}}

\definecolor{c1}{cmyk}{0,0.6175,0.8848,0.1490}
\definecolor{c2}{cmyk}{0.1127,0.6690,0,0.4431}
\definecolor{c3}{cmyk}{0.3081,0,0.7209,0.3255}
\definecolor{c4}{cmyk}{0.6765,0.2017,0,0.0667}
\definecolor{c5}{cmyk}{0,0.8765,0.7099,0.3647}

\definecolor{lightgrey}{rgb}{0.93,0.93,0.93}

\newtcbox{\hlprimarytab}{on line, rounded corners, box align=base, colback=c3!10,colframe=white,size=fbox,arc=3pt, before upper=\strut, top=-2pt, bottom=-4pt, left=-2pt, right=-2pt, boxrule=0pt}
\newtcbox{\hlsecondarytab}{on line, box align=base, colback=red!10,colframe=white,size=fbox,arc=3pt, before upper=\strut, top=-2pt, bottom=-4pt, left=-2pt, right=-2pt, boxrule=0pt}
\newcommand{\dashifted}{\raisebox{0.5\depth}{\tiny$\downarrow$}}
\newcommand{\uashifted}{\raisebox{0.5\depth}{\tiny$\uparrow$}}
\newcommand{\da}[1]{{\scriptsize\hlsecondarytab{\dashifted{#1}}}}
\newcommand{\ua}[1]{{\scriptsize\hlprimarytab{\uashifted{#1}}}}

\title{Explainable Fault Localization for Programming Assignments via LLM-Guided Annotation}

\author{\IEEEauthorblockN{Fang Liu\IEEEauthorrefmark{1}, Tianze Wang\IEEEauthorrefmark{1}, Li Zhang, Zheyu Yang, Jing Jiang\IEEEauthorrefmark{2}, Zian Sun}
\IEEEauthorblockA{State Key Laboratory of Complex \& Critical Software Environment, School of Computer Science and Engineering \\ Beihang University, Beijing, China
    \\ Email: \{fangliu, wangtz, lily, jiangjing, sza\}@buaa.edu.cn, yangzheyu00@hotmail.com}
\IEEEcompsocitemizethanks{\IEEEcompsocthanksitem\IEEEauthorrefmark{1}Fang Liu and Tianze Wang contributed equally to this paper.}
\IEEEcompsocitemizethanks{\IEEEcompsocthanksitem\IEEEauthorrefmark{2}Corresponding author.}
}

\maketitle

\begin{abstract}
Providing timely and personalized guidance for students' programming assignments, particularly by indicating fine-grained error locations with explanations, offers significant practical value for helping students complete assignments and enhance their learning outcomes. In recent years, various automated Fault Localization (FL) techniques, particularly those leveraging Large Language Models (LLMs), have demonstrated promising results in identifying errors in programs. However, existing fault localization techniques face challenges when applied to educational contexts. Most approaches operate at the method level without explanatory feedback, resulting in granularity too coarse for students who need actionable insights to identify and fix their errors. While some approaches attempt line-level fault localization, they often depend on predicting line numbers directly in numerical form, which is ill-suited to LLMs.
To address these challenges, we propose \MethodName{}, a fine-grained, explainable \underline{F}ault \underline{L}ocalization method tailored for programming assignments via LLM-guided \underline{A}nnotation and \underline{M}odel \underline{E}nsemble. \MethodName{} leverages rich contextual information specific to programming assignments to guide LLMs in identifying faulty code lines. Instead of directly predicting line numbers, we prompt the LLM to annotate faulty code lines with detailed explanations, enhancing both localization accuracy and educational value. To further improve reliability, we introduce a weighted multi-model voting strategy that aggregates results from multiple LLMs to determine the suspiciousness of each code line. Extensive experimental results demonstrate that \MethodName{} outperforms state-of-the-art fault localization baselines on programming assignments, successfully localizing 207 more faults at top-1 over the best-performing baseline. Beyond educational contexts, \MethodName{} also generalizes effectively to general-purpose software codebases, outperforming all baselines on the Defects4J benchmark.
\end{abstract}


\section{Introduction}
Programming assignments are essential for students to develop coding skills. Since manual grading of these assignments is a time-consuming and error-prone process, Online Judge (OJ) systems are widely employed for automated evaluation~\cite{cheang2003automated,douce2005automatic,ala2005survey,ihantola2010review}.
To complete an assignment, students submit their programs to an online judge system, which automatically executes the programs against a suite of predefined tests. If the program fails any test, students are expected to independently identify and fix the errors in their code before resubmitting. However, the feedback from such systems is often overly simplistic---commonly limited to verdicts such as ``Wrong Answer'' or ``Runtime Error''~\cite{wasik2018survey,watanobe2022online}---which poses significant challenges for less skilled students, who may struggle with identifying the errors, let alone fixing them~\cite{singh2013automated}. Some students may even become discouraged and give up on the assignment altogether, putting them at risk of failing the course. Given that instructors are often responsible for hundreds or even thousands of students~\cite{masters2011brief}, it is impractical for them to provide timely and personalized guidance to each individual. 
Therefore, offering automated feedback with more informative insights---particularly by indicating fine-grained error locations with explanations---offers significant practical value for helping students complete assignments and enhance their learning outcomes~\cite{liu2023empirical}.

Automated fault localization techniques offer promising avenues for addressing this challenge~\cite{araujo2016applying,edmison2019experiences,gupta2019neural,nguyen2022ffl}. 
\emph{Fault Localization (FL)} is a critical activity within the software debugging process, aiming to identify the specific program elements (\textit{e.g.}, lines, methods, classes, or files) that cause observed program failures~\cite{wong2016survey}. Over the years, various automated fault localization techniques have been proposed. One prominent paradigm is \emph{Spectrum-Based Fault Localization (SBFL)}~\cite{abreu2006evaluation,abreu2009spectrum,wong2013dstar}, which postulates that program elements executed more frequently by failing tests than by passing tests are more likely to be faulty. 
Another notable paradigm is \emph{Learning-Based Fault Localization (LBFL)}~\cite{b2016learning,gupta2019neural,lou2021boosting,meng2022improving},
which leverages machine learning and deep learning techniques to improve fault localization. Most recently, the rise of Large Language Models (LLMs) has further advanced this field. Several studies have explored fault localization using LLMs, demonstrating promising results~\cite{yang2024large,qin2024agentfl,widyasari2024demystifying,xu2025flexfl}. 

Among the existing approaches, most operate at the method level and are evaluated on the Defects4J~\cite{just2014defects4j} benchmark. For example, 
\citet{xu2025flexfl} present a two-stage LLM-based fault localization framework, FlexFL, which leverages existing fault localization techniques for search space reduction and LLMs for localization refinement. Only a few approaches attempt line-level fault localization. LLMAO~\cite{yang2024large} fine-tunes lightweight adapters on top of LLMs to predict buggy lines. FuseFL~\cite{widyasari2024demystifying} combines SBFL results, test outcomes, and code descriptions to guide LLMs in predicting buggy lines and generating explanations.

However, despite the existence of various fault localization techniques, most are ineffective when applied to educational contexts due to the following challenges:
\ding{182}~\textbf{Coarse localization granularity}: Most existing techniques operate at the method level~\cite{qin2024agentfl,xu2025flexfl}, with relatively few supporting line-level localization. Such coarse granularity is often insufficient for students, who need more actionable insights to identify and fix their errors.
\ding{183}~\textbf{Inaccurate faulty line prediction}: While some techniques attempt line-level fault localization, most of them depend on predicting line numbers directly in numerical form~\cite{wu2023large,widyasari2024demystifying}, which is particularly ill-suited to LLMs. Due to the inherent limitations of LLMs in numerical understanding and processing~\cite{yang2024number}, line numbers generated in this way are frequently inaccurate, making such techniques unreliable in practice.
\ding{184}~\textbf{Lack of explainability}: Most existing techniques simply identify fault locations without providing explanations, limiting their educational value.
\ding{185}~\textbf{Artifact availability gap}: Some techniques depend on software development artifacts such as version control histories or bug reports~\cite{sohn2017fluccs,xu2025flexfl}, which are generally unavailable in educational contexts. Some others utilize documentation or code comments~\cite{kang2024quantitative}, which are often missing, incomplete, or unreliable in student programs due to their highly inconsistent code quality and limited development maturity~\cite{keuning2017code,ostlund2023s}, especially among less skilled students. In contrast, artifacts specific to programming assignments, such as problem statements and reference programs, remain largely underutilized despite their potential to enhance fault localization effectiveness.

In this paper, we propose \MethodName{}, a \emph{fine-grained}, \emph{explainable} fault localization method tailored for programming assignments. 
Specifically, \MethodName{} empowers LLMs to identify faults in programming assignments by leveraging rich contextual information, including problem statements, test outcomes, and reference programs retrieved from historical submissions. To enhance fault localization accuracy and educational value, rather than predicting line numbers directly in numerical form, we prompt the LLM to annotate faulty lines in the program with detailed natural language explanations. To further improve the reliability of localization results, we perform multiple rounds of fault location annotation using different LLMs, and then aggregate the results to determine the suspiciousness of each line via a weighted voting strategy.

To assess the effectiveness of \MethodName{} on programming assignments, we introduce \DatasetName{}, a dataset encompassing both single-file and project-level submissions from two programming courses. Experimental results show that \MethodName{} outperforms previous state-of-the-art fault localization baselines on \DatasetName{}, localizing 207 more faults at top-1 over the best-performing baseline \cite{wu2023large}, demonstrating its effectiveness in educational contexts.
We further evaluate \MethodName{} on the Defects4J~\cite{just2014defects4j} benchmark, where it achieves state-of-the-art performance, confirming its generalization capability to general-purpose software codebases beyond educational contexts. Additionally, experiments reveal that \MethodName{}'s fault localization results can provide valuable guidance for automated program repair.

Our main contributions are as follows:
\begin{itemize}
    \item We introduce \DatasetName{}, a new dataset consisting of 1,932 faulty-fixed pairs of student submissions from two programming courses collected from an online judge system, encompassing both single-file and project-level submissions.
    \item We propose \MethodName{}, a fine-grained, explainable fault localization method tailored for programming assignments, which supports line-level fault localization with detailed natural language explanation for each identified faulty line.
    \item We conduct a comprehensive evaluation of \MethodName{}, demonstrating its superior performance compared to state-of-the-art fault localization methods across both programming assignments and general-purpose software codebases. The results highlight \MethodName{}'s strong generalization capability and practical applicability. To facilitate further research, the code and data are available at \url{https://github.com/FLAME-FL/FLAME}.
\end{itemize}

\begin{figure*}[t]
    \centering
    \setlength{\abovecaptionskip}{0.1cm}
    \includegraphics[width=\linewidth]{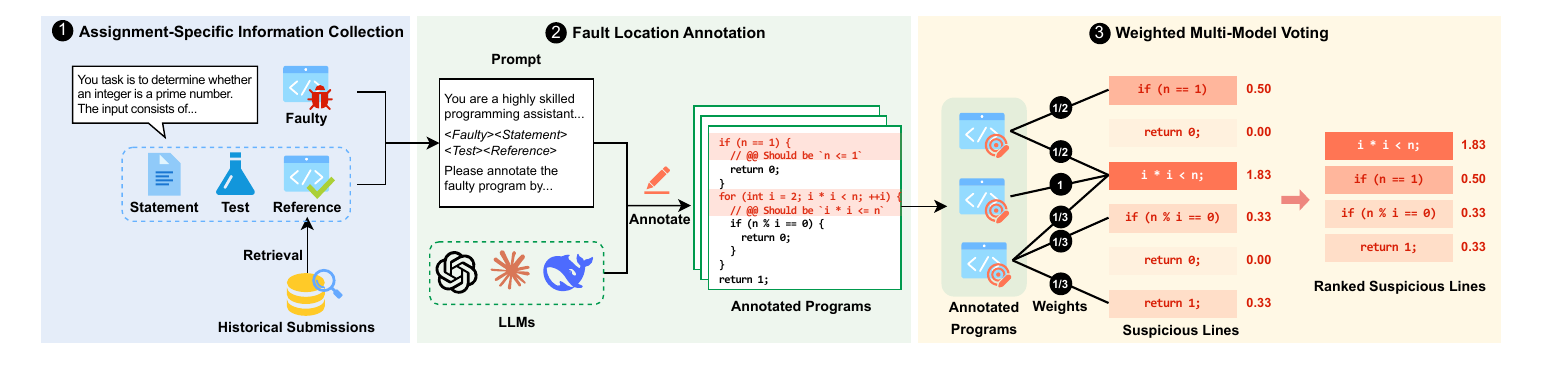}
    \caption{Overview of \MethodName{}, including \ding{182} Assignment-Specific Information Collection, \ding{183} Fault Location Annotation, and \ding{184} Weighted Multi-Model Voting.
    }
    \label{fig:Overview}
    \vspace{-0.3cm}
\end{figure*}

\section{Related Work}

\subsection{Fault Localization Techniques for Programming Assignments}

Several fault localization techniques are designed for programming assignments.
\citet{araujo2016applying} conducted an empirical study on the application of SBFL to novice programs, highlighting its effectiveness while also showing its limitations in certain scenarios.
NBL~\cite{gupta2019neural} combined convolutional neural networks with prediction attribution techniques to localize buggy lines in student programs.
FFL~\cite{nguyen2022ffl} used graph-based syntax and semantic reasoning with graph neural networks to localize buggy statements in student programs.
FuseFL~\cite{widyasari2024demystifying} combined SBFL results, test results, and code descriptions to guide LLMs in localizing the fault and generating step-by-step reasoning about the fault.
These techniques primarily focus on single-file submissions with no more than a few hundred lines of code, and may not scale effectively to project-level submissions with thousands of lines of code. Moreover, the two learning-based techniques (NBL~\cite{gupta2019neural} and FFL~\cite{nguyen2022ffl}) necessitate effort in preparing labeled datasets for supervised training, which hinders their broader adoption.

\subsection{Fault Localization Techniques for General-Purpose Software Codebases} 
Most existing fault localization techniques are designed for general-purpose software codebases. Among these, Learning-Based Fault Localization (LBFL), especially deep learning-based techniques, has gained prominence over the years.
CNN-FL~\cite{zhang2019cnn} and DeepRL4FL~\cite{li2021fault} treated fault localization as a pattern recognition task, using convolutional neural networks to learn discriminative representations of code coverage matrices. DeepFL~\cite{li2019deepfl} incorporated various suspiciousness-value-based, fault-proneness-based, and textual-similarity-based features to train deep models. GRACE~\cite{lou2021boosting} used graph neural networks to learn from graph-based coverage representations, enabling better utilization of coverage information. 
LLMAO~\cite{yang2024large} fine-tuned lightweight bidirectional adapters on top of pre-trained left-to-right language models to directly predict faulty lines without relying on test coverage information. Due to the nature of the learning algorithms used in these techniques, their fault localization results are not explainable.

Most recently, the rise of Large Language Models (LLMs) has further advanced this field. Several studies have explored fault localization using LLMs, demonstrating promising results. \citet{wu2023large} conducted a comprehensive evaluation of LLMs (specifically ChatGPT-3.5 and ChatGPT-4) on fault localization, examining their sensitivity to prompt design and context granularity. AgentFL~\cite{qin2024agentfl} scaled LLM-based fault localization to project-level contexts using a multi-agent system that simulates human debugging behavior. AutoFL~\cite{kang2024quantitative} equipped LLMs with function-calling capabilities to autonomously navigate large repositories and generate explanations along with suggested fault locations. FlexFL~\cite{xu2025flexfl} enabled flexible and effective fault localization through a two-stage LLM-based framework leveraging existing fault localization techniques for search space reduction and open-source LLMs for localization refinement. All these techniques operate at the method-level, and none of them address the explainability of fault localization except AutoFL~\cite{kang2024quantitative}.

To sum up, most existing techniques are ineffective when applied to educational contexts due to their coarse localization granularity and lack of explainability.
Moreover, these techniques are primarily designed for general-purpose software codebases and do not take advantage of artifacts specific to programming assignments, further limiting their effectiveness.

\section{Methodology}

We propose \MethodName{}, a \emph{fine-grained}, \emph{explainable} fault localization method for programming assignments based on large language models. An overview of \MethodName{} is illustrated in \autoref{fig:Overview}. 
Given a faulty program $P=(\ell_1,\ell_2,\dots,\ell_n)$, where $\ell_i$ is the $i$-th line of the program, \MethodName{} begins by collecting assignment-specific auxiliary information $I$, such as the problem statement, test outcomes, and reference program. Then both the faulty program $P$ and the assignment-specific auxiliary information $I$ are used to prompt the LLM to determine the suspiciousness of each line in $P$, finally producing a ranked sequence of suspicious lines $S=(\ell'_1,\ell'_2,\dots,\ell'_n)$ with our proposed weighted voting strategy, ordered by their suspiciousness scores $s(\ell)$ in descending order.
Specifically, our method consists of three stages:
\begin{itemize}[leftmargin=*]
    \item \textbf{Assignment-Specific Information Collection}:
    To provide critical hints for understanding program semantics and narrow down potential fault candidates, we collect assignment-specific auxiliary information such as problem statements, test outcomes, and reference programs, and incorporate them into the prompt to enhance LLM's fault localization effectiveness.
    \item \textbf{Fault Location Annotation}: Instead of predicting line numbers directly in numerical form, we prompt the LLM to annotate faulty lines in the program with detailed natural language explanations, ensuring both localization accuracy and educational value.
    \item \textbf{Weighted Multi-Model Voting}: To mitigate single-model biases and improve reliability, we perform multiple rounds of fault location annotation using different LLMs, then aggregate the results via a weighted voting strategy that prioritizes consensus predictions, producing a more robust final ranking of suspicious lines.
\end{itemize}
Below we present the details of each stage.

\subsection{Assignment-Specific Information Collection}

We collect the following assignment-specific auxiliary information $I$ to enhance LLM's fault localization effectiveness, which provides critical insights into program semantics and narrows the scope of potential fault candidates. 

\subsubsection{Problem Statement}
\begin{figure}[h]
    \centering
    \setlength{\abovecaptionskip}{0.1cm}
    \includegraphics[scale=1]{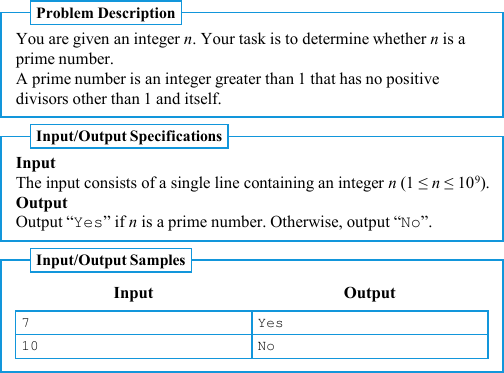}
    \caption{An example of a problem statement, consisting of the problem description, input/output specifications, and input/output samples.}
    \label{fig:Statement}
    \vspace{-0.3cm}
\end{figure}

The \emph{statement} of a problem comprehensively and clearly defines the expected behavior of a program, which is valuable for fault localization. As shown in \autoref{fig:Statement}, it consists of \emph{problem description}, \emph{input/output specifications}, and \emph{input/output samples}:
\begin{itemize}[leftmargin=*]
    \item The \emph{problem description} defines the problem that needs to be solved.
    \item The \emph{input/output specifications} define the expected input and output format.
    \item The \emph{input/output samples} provide one or more sample inputs and their expected outputs, possibly with explanations, aiding in understanding the problem.
\end{itemize}

\subsubsection{Test Outcomes}

Online judge systems evaluate programs by executing them against a suite of predefined \emph{tests}. For each test, the system issues a \emph{verdict}, which is typically one of \emph{Accepted (AC)}, \emph{Wrong Answer (WA)}, \emph{Runtime Error (RE)}, \emph{Time Limit Exceeded (TLE)}, and \emph{Memory Limit Exceeded (MLE)}. A program is accepted if and only if it receives AC for all tests. If a program is rejected, it must have received a non-AC verdict on at least one test (or, in some cases, failed to compile and thus could not be tested). We utilize these failing tests as follows:
\begin{itemize}[leftmargin=*]
    \item If the program fails to compile, we use the diagnostic messages from the compiler.
    \item Otherwise, we extract the input, expected output, and verdict of one failing test, prioritizing tests with a verdict of WA, followed by RE, TLE, and MLE. If the verdict is WA, the actual output is also included.
\end{itemize}

\subsubsection{Reference Programs}

Online judge systems typically maintain a repository of historical submissions. If a previously accepted submission to the current problem exhibits similarity to the faulty program, it may offer critical guidance for fault localization. To identify such submissions, we use a text embedding model to generate embeddings for the faulty program and all accepted historical submissions to the current problem. The most semantically similar submission based on cosine similarity is then retrieved as the reference program.

\subsection{Fault Location Annotation}

Next, we prompt an LLM to annotate faulty lines based on the faulty program $P=(\ell_1,\ell_2,\dots,\ell_n)$ and the collected auxiliary information $I$. Ideally, the LLM should directly output the line numbers of faulty lines, possibly with explanations, as done in prior work~\cite{wu2023large,widyasari2024demystifying}. However, while identifying line numbers is trivial for humans, LLMs struggle with this task due to their inherent limitations in numerical understanding and processing~\cite{yang2024number}. As a result, line number predictions with LLMs are frequently inaccurate, making such techniques unreliable in practice. Our preliminary experiments further confirm that directly prompting the LLM to output line numbers often yields suboptimal results, even when line numbers are explicitly prepended to each line in the input faulty program.

\begin{figure}[t]
    \centering
    \setlength{\abovecaptionskip}{0.1cm}
    \includegraphics[scale=0.7]{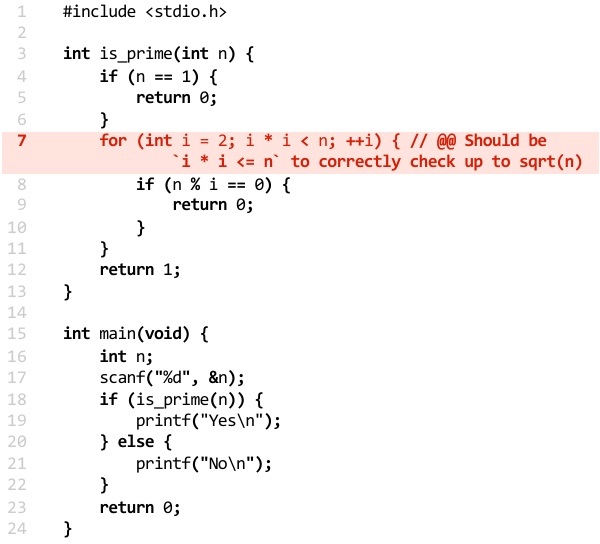}
    \caption{An example of an annotated program. Line 7 is marked as faulty by appending a marker \texttt{// @@} at the end of the line, followed by an explanation stating that \texttt{i * i < n} should be corrected to \texttt{i * i <= n}.}
    \label{fig:AnnotatedProgram}
    \vspace{-0.3cm}
\end{figure}

To address this, instead of predicting line numbers directly, we prompt the LLM to return an \emph{annotated} version of the program, in which each faulty line is marked by appending a marker \texttt{// @@} at the end, followed by a detailed natural language explanation, resulting in a set of annotated lines $A\subseteq\{\ell_1,\ell_2,\dots,\ell_n\}$ and their explanations $e(\ell)$ ($\ell\in A$). An example is shown in \autoref{fig:AnnotatedProgram}.

To accurately identify the annotated lines and extract their explanations, we perform fuzzy matching between the lines in the annotated and original programs, instead of relying solely on the positions of the markers, which can be misaligned due to code formatting changes introduced by the LLM, such as inserting blank lines, adjusting indentation, or adding comments.  Specifically, we consider two lines to match if the cosine similarity of their vector representations (computed using the \texttt{HashingVectorizer} from the \emph{scikit-learn}~\cite{scikit-learn} library) exceeds a threshold of $0.9$, after stripping out any appended annotations. To align an annotated line with its original counterpart, we perform a bidirectional linear search (upward and downward) starting from the line with the same line number, stopping once a match is found.

For project-level submissions from advanced courses (such as software engineering, compiler technology, and operating systems), where multiple source files are typically involved, we concatenate the file paths and their contents into a single plain-text document, which serves as the faulty program to be annotated. However, unlike single-file submissions, project-level submissions are often much longer and may exceed the output token limits of proprietary inference services, which are typically lower than the input token limits. To mitigate this, we instruct the LLM to collapse non-essential code sections into \colorbox{lightgray}{$\cdots$} when deemed appropriate, thereby reducing the number of output tokens while preserving critical contextual information for analysis.

\subsection{Weighted Multi-Model Voting}

A single annotation result generated by a single model may be suboptimal due to inherent model biases and the stochastic nature of sampling. Different models exhibit varying strengths and weaknesses stemming from differences in model architecture, parameter size, training data, training methodology, \textit{etc}. LLM ensemble has shown promise in leveraging the individual strengths of diverse models through strategic combination~\cite{chen2025harnessing}. Additionally, techniques such as Self-Consistency~\cite{wang2022self} have shown that aggregating multiple generations from a single model can also improve performance.
Motivated by this, we employ $m$ state-of-the-art LLMs, each performing $r$ rounds of fault location annotation. We then aggregate the annotation results to determine the suspiciousness of each line via a weighted voting strategy inspired by prior work~\cite{kang2024quantitative}. Specifically, given a faulty program $P=(\ell_1,\ell_2,\dots,\ell_n)$, we formalize the process as follows:

\noindent\textbf{\textit{(1) Model Selection and Annotation}}
\begin{itemize}
    \item \textbf{Step 1:} Select $m$ state-of-the-art LLMs.
    \item \textbf{Step 2:} For each LLM, perform $r$ rounds of fault location annotation, resulting in a total of $m\cdot r$ annotations:
    \begin{equation}
        \mathscr{A}=\{A_1,A_2,\dots,A_{m\cdot r}\}.
    \end{equation}
\end{itemize}

\noindent\textbf{\textit{(2) Weight Assignment}}
\begin{itemize}
    \item \textbf{Step 3:} For each annotation $A$, we assign a normalized weight to each annotated line $\ell$:
    \begin{equation}
        w_A(\ell)=\frac{1}{|A|}\quad(\ell\in A),
    \end{equation}
    where $|A|$ is the number of annotated lines in $A$. For lines not annotated, we assign a weight of $0$:
    \begin{equation}
        w_A(\ell)=0\quad(\ell\notin A).
    \end{equation}
\end{itemize}

\noindent\textbf{\textit{(3) Suspiciousness Score Computation and Ranking}}
\begin{itemize}
    \item \textbf{Step 4:} Compute the suspiciousness score of each line $\ell$ in $P$ by summing the weights across all annotations:
    \begin{equation}
        s(\ell)=\sum_{A\in\mathscr{A}}w_A(\ell).
    \end{equation}
    \item \textbf{Step 5:} Rank suspicious lines in descending order of their suspiciousness scores $s(\ell)$ (breaking ties by ascending line numbers) to get the final ranked sequence:
    \begin{equation}
        S=(\ell'_1,\ell'_2,\dots,\ell'_n),
    \end{equation}
    where $S$ is a permutation of $(\ell_1,\ell_2,\dots,\ell_n)$, satisfying
    \begin{equation}
        s(\ell'_1)\ge s(\ell'_2)\ge\cdots\ge s(\ell'_n).
    \end{equation}
\end{itemize}

For each suspicious line $\ell$ with a non-zero suspiciousness score $s(\ell)$, we choose the explanation generated by the model that assigned the highest weight $w_A(\ell)$ to the line as the final explanation. We discard explanations generated by other models to avoid overwhelming the user with excessive information.

\section{Experimental Setup}

\subsection{Research Questions}

We aim to answer the following research questions in our evaluation:
\begin{itemize}[leftmargin=*]
    \item \textbf{RQ1: How effective is \MethodName{} compared to baseline methods?} This question evaluates the performance of \MethodName{} on programming assignments and compare it against state-of-the-art fault localization techniques.
    \item \textbf{RQ2: How effective are the design choices of \MethodName{}?} This question examines the contribution of each component within \MethodName{} to overall performance.
    \item \textbf{RQ3: How well does \MethodName{} generalize to general-purpose software codebases?} This question assesses the generalization capability of \MethodName{} to general-purpose software codebases beyond educational contexts by evaluating its performance on the Defects4J~\cite{just2014defects4j} benchmark.
    \item \textbf{RQ4: How helpful are the results of \MethodName{} in assisting automated program repair?} This question investigates whether \MethodName{}'s fault localization results can enhance the bug-fixing process when integrated with APR tools.
\end{itemize}

\subsection{Datasets}
To evaluate the effectiveness of \MethodName{} in fault localization for programming assignments on both single-file and project-level submissions, we introduce \DatasetName{}, a new dataset consisting of student submissions from two programming courses, \emph{Data Structures} and \emph{Compiler Technology}, collected from the online judge system of Beihang University.
The detailed information of \DatasetName{} is as follows:

\textbf{Data Structures (\DatasetName{}\textsubscript{DS})}: We collect student submissions from the \emph{Data Structures} course, a foundational course for second-year undergraduate computer science students. In this course, students are tasked with solving programming problems involving various data structures. Each submission is a single-file C program. We select 10 programming problems covering a diverse range of topics, including linked lists, stacks, queues, trees, graphs, and strings. Submissions to these problems typically contain 50--200 lines of code. For each problem, we randomly select 150 students from all students enrolled in the 2024 academic year who made at least one failing attempt before passing all tests. For each student, we pair their final accepted submission with their last rejected submission. This results in 1,500 submission pairs in total. Additionally, we collect submissions to the same problems from the 2023 academic year to serve as reference programs.

\textbf{Compiler Technology (\DatasetName{}\textsubscript{CT})}: We also collect student submissions from the \emph{Compiler Technology} course, a specialized course for third-year undergraduate computer science students. In this course, students are tasked with building a compiler that translates source programs into assembly. The development process is divided into several incremental stages, such as lexical analysis, syntax analysis, semantic analysis, and code generation. Each stage is formulated as a programming problem, with a statement and a suite of tests. Different from Data Structures, each submission in Compiler Technology is a full C++ or Java project consisting of multiple source files. 
Our study focuses on the first two critical stages, ``lexical analysis'' and ``syntax analysis.'' Submissions to ``lexical analysis'' and ``syntax analysis'' typically contain about 500 and 2,000 lines of code, respectively. For each stage, we select all students enrolled in the 2024 academic year who made at least one failing attempt before passing all tests. 
For each student, we pair their final accepted submission with their last rejected submission. 
We exclude submission pairs involving file creation or deletion between the two submissions. 
This results in 239 submission pairs for ``lexical analysis'' and 193 submission pairs for ``syntax analysis.'' Due to more complex requirements and flexible implementations, obtaining similar reference programs in Compiler Technology is challenging. Therefore, we exclude reference programs from this dataset. 

The detailed statistics of \DatasetName{} are summarized in \autoref{tab:DatasetStatistics}. For all submission pairs, following prior work~\cite{tan2017codeflaws,yang2024large}, we treat lines that differ between the two submissions as the ground truth of faulty lines.

\begin{table}[t]
    \setlength{\abovecaptionskip}{0.1cm}
    \caption{Statistics of \DatasetName{}, including the number of submission pairs (\# Submission Pairs) in each subset, the programming language of the submissions (PL), the number of source files per submission (\# Source Files), and the number of lines of code per submission (\# LoC).}
    \centering
    \resizebox{\linewidth}{!}{\begin{tabular}{lcccccc}
        \toprule
        \multirow{2}{*}{\textbf{Dataset}} & \multirow{2}{*}{\# \textbf{Submission Pairs}} & \multirow{2}{*}{\textbf{PL}} & \multicolumn{2}{c}{\textbf{\# Source Files}} & \multicolumn{2}{c}{\# \textbf{LoC}}\\
        \cmidrule(r){4-5}
        \cmidrule(r){6-7}
        & & & Avg. & Med. & Avg. & Med.\\
        \midrule
        \DatasetName{}\textsubscript{DS} & 1,500 & C & 1 & 1 & 72 & 61\\
        \DatasetName{}\textsubscript{CT} & 432 & C++, Java & 20 & 7 & 1,371 & 620\\
        \bottomrule
    \end{tabular}
    }
    \label{tab:DatasetStatistics}
    \vspace{-0.3cm}
\end{table}

Moreover, to assess the generalization capability of our method to general-purpose software codebases, we also evaluate it on \emph{Defects4J}~\cite{just2014defects4j}, a widely used dataset containing real bugs from real-world open-source projects. Following the empirical study by \citet{wu2023large}, we use a representative subset of Defects4J v1.2.0 containing 408 bugs for our evaluation.

\subsection{Metrics}

We adopt the \emph{top-$k$} metric to evaluate the effectiveness of \MethodName{} and baselines, which is widely adopted in fault localization tasks~\cite{qin2024agentfl,widyasari2024demystifying,xu2025flexfl}. This metric is defined as the number of successfully localized programs, where a localization is considered successful if the faulty line appears within the top $k$ most suspicious lines. In scenarios where a program contains multiple faulty lines, a localization is considered successful if at least one of the faulty lines appears within the top-$k$ most suspicious lines, which is consistent with established practice~\cite{widyasari2024demystifying}.
Moreover, we also report the \textit{top-$k$ accuracy}, defined as the proportion of successfully localized programs across all programs.

\subsection{Baseline Methods}

We compare \MethodName{} with the following baseline methods, including several state-of-the-art LLM-based fault localization methods and a spectrum-based fault localization method\footnote{We also compare \MethodName{} with single-model baselines (\textit{e.g.}, Claude-only). For convenience of presentation, these baselines are treated as variants of \MethodName{} and are discussed in RQ2.}:
\begin{itemize}[leftmargin=*]
    \item \textbf{LLMAO}~\cite{yang2024large}: A learning-based fault localization method that fine-tunes bidirectional adapters on top of pre-trained models to predict faulty lines. We fine-tune it using submissions from the 2023 academic year of the Data Structures course for evaluation on \DatasetName{}\textsubscript{DS}. We exclude it from evaluation on \DatasetName{}\textsubscript{CT} due to its insufficient input length limit, which is shorter than most programs in \DatasetName{}\textsubscript{CT}.
    \item \textbf{FuseFL}~\cite{widyasari2024demystifying}: An LLM-based fault localization method that combines SBFL results, test results, and code descriptions to guide LLMs in localizing the fault and generating step-by-step reasoning about the fault.
    This method was originally implemented using GPT-3.5~\cite{brown2020language} via the ``September 11, 2023'' release of ChatGPT~\cite{chatgpt}, which is no longer accessible. As a replacement, we implement it using the more advanced GPT-4.1~\cite{openai2025gpt41} (\texttt{gpt-4.1-2025-04-14}) in our evaluation.
    \item \textbf{ChatGPT-4 (Log)}~\cite{wu2023large}: A method used in the empirical study by \citet{wu2023large} that directly prompts ChatGPT-4 with the faulty program and basic instructions, followed by a failing test and its corresponding error log.
    This method was originally implemented using GPT-4~\cite{openai2024gpt4} via the ChatGPT Web interface~\cite{chatgpt}, which is no longer accessible. As a replacement, we implement it using the more advanced GPT-4.1~\cite{openai2025gpt41} (\texttt{gpt-4.1-2025-04-14}) in our evaluation.
    \item \textbf{Ochiai}~\cite{abreu2006evaluation}: The commonly used spectrum-based fault localization method using the Ochiai coefficient.
\end{itemize}

\subsection{Implementation Details}

For reference program retrieval, we use OpenAI's \texttt{text-embedding-3-small}~\cite{openai2024embedding} model to generate program embeddings. Regarding the weighted multi-model voting stage, we employ $m=3$ state-of-the-art LLMs, \textit{i.e.}, Claude 3.7 Sonnet~\cite{anthropic2025claude}, 
DeepSeek-V3~\cite{deepseekai2025deepseekv3}, and DeepSeek-R1~\cite{deepseekai2025deepseekr1}, all of which have demonstrated strong performance in code-related tasks \cite{anthropic2025claude,deepseekai2025deepseekv3,deepseekai2025deepseekr1,aider,livebench}, and our preliminary experiments also show that using these models strikes a good balance between performance and inference cost. 
Each model performs $r=2$ rounds of fault location annotation for voting. 
For all models, we set the temperature to $0.1$ after evaluating several candidate values in preliminary experiments, which can better balance output diversity and reliability.

\begin{table*}[t]
    \centering
    \setlength{\abovecaptionskip}{0.1cm}
    \caption{Fault localization results on \DatasetName{}. The number inside the parentheses is the \textit{top-$k$ accuracy} computed within each subset.}
    \begin{tabular}{llcccc}
        \toprule
        \textbf{Dataset} & \textbf{Method} & \textbf{Top-1} & \textbf{Top-3} & \textbf{Top-5} & \textbf{Top-10}\\
        \midrule
        \multirow{5}{*}{\DatasetName{}\textsubscript{DS}} & LLMAO & 232 (15.5\%) & 471 (31.4\%) & 644 (42.9\%) & 858 (57.2\%)\\
        & FuseFL & 363 (24.2\%) & 515 (34.3\%) & 579 (38.6\%) & 635 (42.3\%)\\
        & ChatGPT-4 (Log) & 792 (52.8\%) & 1054 (70.3\%) & 1141 (76.1\%) & 1183 (78.9\%)\\
        & Ochiai & 159 (10.6\%) & 305 (20.3\%) & 457 (30.5\%) & 705 (47.0\%)\\
        & \textbf{\MethodName{}} & \textbf{941 (62.7\%)} & \textbf{1195 (79.7\%)} & \textbf{1269 (84.6\%)} & \textbf{1309 (87.3\%)}\\
        \midrule
        \multirow{4}{*}{\DatasetName{}\textsubscript{CT}} & FuseFL & 87 (20.1\%) & 124 (28.7\%) & 150 (34.7\%) & 150 (34.7\%) \\
        & ChatGPT-4 (Log) & 125 (28.9\%) & 205 (47.5\%) & 218 (50.5\%) & 222 (51.4\%)\\
        & Ochiai & 68 (15.7\%) & 117 (27.1\%) & 146 (33.8\%) & 177 (41.0\%)\\
        & \textbf{\MethodName{}} & \textbf{183 (42.4\%)} & \textbf{259 (60.0\%)} & \textbf{285 (66.0\%)} & \textbf{292 (67.6\%)}\\
        \bottomrule
    \end{tabular}
    \label{tab:RQ1}
    \vspace{-0.3cm}
\end{table*}

\section{Results and Analysis}

\subsection{RQ1: How effective is \MethodName{} compared to baseline methods?}

To address this research question, we evaluate \MethodName{} against four baseline methods, \textit{i.e.}, LLMAO~\cite{yang2024large}, FuseFL~\cite{widyasari2024demystifying}, ChatGPT-4 (Log)~\cite{wu2023large}, and Ochiai~\cite{abreu2006evaluation}, on \DatasetName{}.
The results are shown in \autoref{tab:RQ1}. 
Overall, \MethodName{} outperforms all baselines, localizing 207 more faults at top-1 over the best-performing baseline (ChatGPT-4 (Log) \cite{wu2023large}), demonstrating its effectiveness.

On \DatasetName{}\textsubscript{DS}, \MethodName{} consistently outperforms all baselines across all top-$k$ metrics, especially in top-1. Specifically, \MethodName{} achieves a top-1 accuracy of 62.7\%, substantially higher than the best-performing baseline, ChatGPT-4 (Log), which achieves only 52.8\%.
Since a substantial number of the faulty programs in the dataset fail all tests, Ochiai, which is based on SBFL, becomes ineffective due to its dependence on both passing and failing tests. This also explains the poor performance of FuseFL, which takes the SBFL results as an important input, and may be misled when those results are unreliable. 
The suboptimal performance of LLMAO can be attributed to the diverse error patterns across student programs. These patterns may not be well represented in the fine-tuning dataset and pose significant challenges for model learning during fine-tuning, resulting in limited generalization capability.
While ChatGPT-4 (Log) outperforms other baselines, its performance remains inferior to \MethodName{}. This gap may stem from the inherent biases of a single model, which could consistently overlook certain error patterns.

Even on the more challenging \DatasetName{}\textsubscript{CT}, \MethodName{} still outperforms all baselines, achieving a top-1 accuracy of 42.4\%. These results demonstrate \MethodName{}'s effectiveness on both single-file and project-level submissions.

\begin{figure}[h]
    \centering
    \setlength{\abovecaptionskip}{0.1cm}
    \includegraphics[scale=0.35]{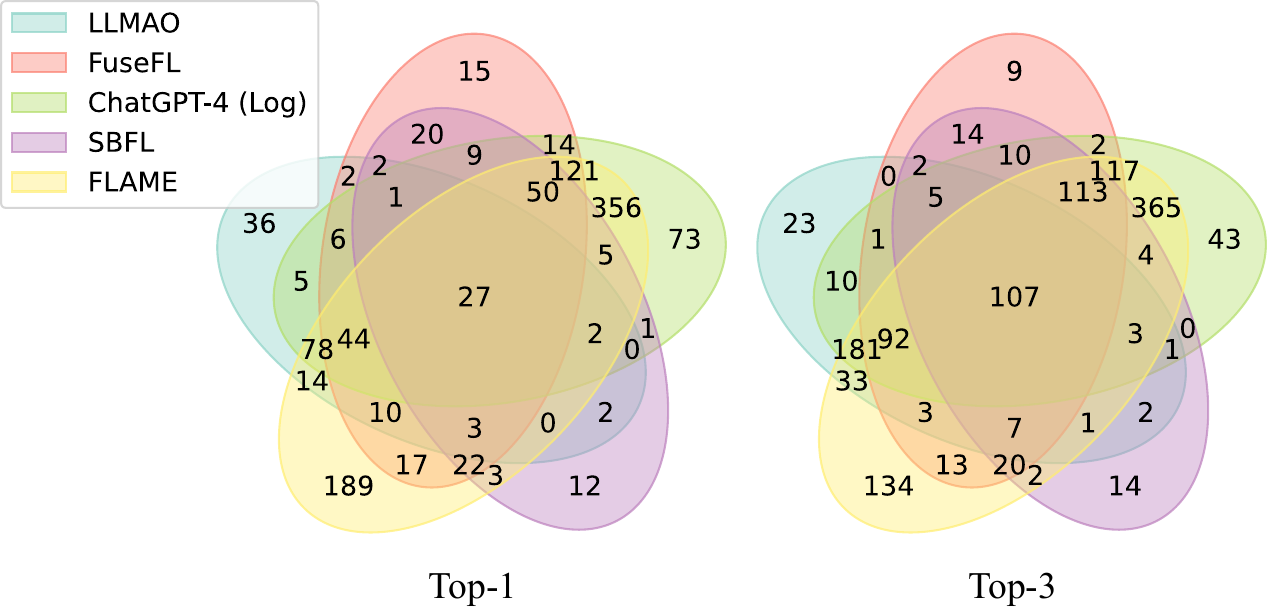}
    \caption{Distribution of localized faults on \DatasetName{}\textsubscript{DS}.}
    \label{fig:RQ1_Venn_DS}
    \vspace{-0.3cm}
\end{figure}

\begin{figure}[h]
    \centering
    \setlength{\abovecaptionskip}{0.1cm}
    \includegraphics[scale=0.35]{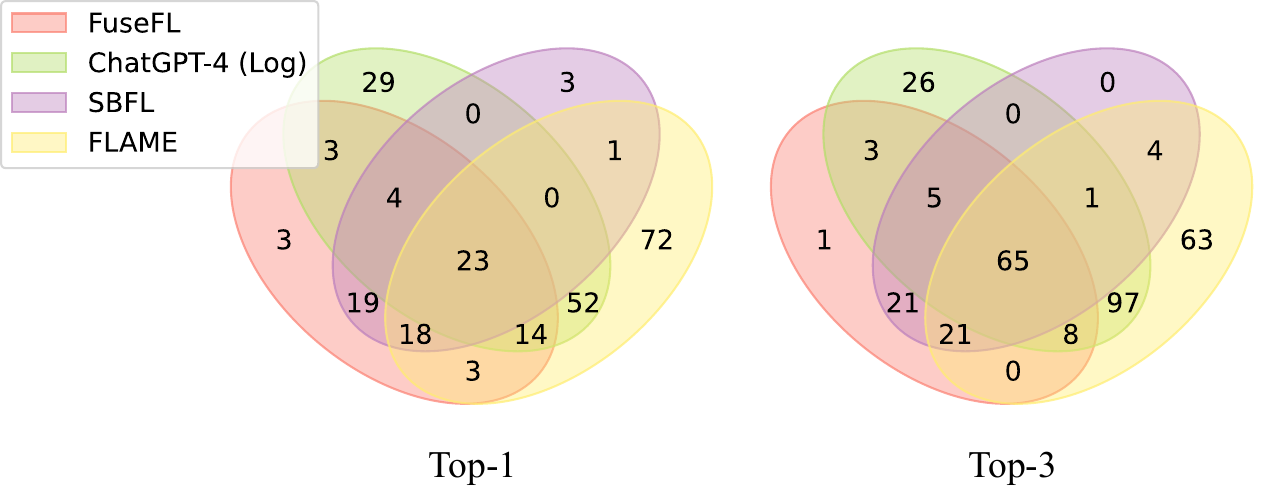}
    \caption{Distribution of localized faults on \DatasetName{}\textsubscript{CT}.}
    \label{fig:RQ1_Venn_CT}
    \vspace{-0.3cm}
\end{figure}

\autoref{fig:RQ1_Venn_DS} and \autoref{fig:RQ1_Venn_CT} further illustrate the set of commonly and uniquely identified faults within top-1 and top-3 by different methods. As seen from the results, \MethodName{} uniquely identified 189 and 72 faults within top-1 on \DatasetName{}\textsubscript{DS} and \MethodName{}\textsubscript{CT}, respectively, substantially outperforming other methods. 
This indicates that \MethodName{} complements existing techniques by detecting error patterns overlooked by other methods, thereby improving overall performance.

\vspace{1mm}
\begin{custommdframed}
\textbf{Answer to RQ1:} \MethodName{} consistently outperforms all baseline methods across all top-$k$ metrics on both datasets, demonstrating its effectiveness in fault localization on both single-file and project-level submissions.
\end{custommdframed}
\vspace{1mm}

\begin{table*}[t]
    \centering
    \setlength{\abovecaptionskip}{0.1cm}
    \caption{Ablation study results of \MethodName{} on \DatasetName{}.}
    \begin{tabular}{lllccccc}
        \toprule
        \textbf{Dataset} & \textbf{Stage} &\textbf{Method} & \textbf{Top-1} & \textbf{Top-3} & \textbf{Top-5} & \textbf{Top-10}\\
        \midrule
        \multirow{9}{*}{\DatasetName{}\textsubscript{DS}} & -  & \textbf{\MethodName{}} & \textbf{941 (62.7\%)} & \textbf{1195 (79.7\%)} & \textbf{1269 (84.6\%)} & \textbf{1309 (87.3\%)}\\
        \cmidrule(lr){2-7}
        & \multirow{3}{*}{Collection} & \MethodName{}\textsubscript{w/o stmt} & 857 (57.1\%) \da{5.6\%} & 1098 (73.2\%) \da{6.5\%} & 1160 (77.3\%) \da{7.3\%} & 1079 (71.9\%) \da{15.4\%}\\
        & & \MethodName{}\textsubscript{w/o test} & 715 (47.7\%) \da{15.0\%} & 1046 (69.7\%) \da{10.0\%} & 1161 (77.4\%) \da{7.2\%} & 1232 (82.1\%) \da{5.2\%}\\
        & & \MethodName{}\textsubscript{w/o ref} & 882 (58.8\%) \da{3.9\%} & 1128 (75.2\%) \da{4.5\%} & 1188 (79.2\%) \da{5.4\%} & 1211 (80.7\%) \da{6.6\%}\\
        \cmidrule(lr){2-7}
        & Annotation & \MethodName{}\textsubscript{num} & 237 (15.8\%) \da{\textbf{46.9\%}} & 453 (30.2\%) \da{\textbf{49.5\%}} & 589 (39.3\%) \da{\textbf{45.3\%}} & 779 (51.9\%) \da{\textbf{35.4\%}}\\
        \cmidrule(lr){2-7}
        & \multirow{4}{*}{Voting} & \MethodName{}\textsubscript{Claude} & 605 (40.3\%) \da{22.4\%} & 1021 (68.1\%) \da{11.6\%} & 1101 (73.4\%) \da{11.2\%} & 1124 (74.9\%) \da{12.4\%}\\
        & & \MethodName{}\textsubscript{DS-V3} & 570 (38.0\%) \da{24.7\%} & 915 (61.0\%) \da{18.7\%} & 972 (64.8\%) \da{19.8\%} & 983 (65.5\%) \da{21.8\%}\\
        & & \MethodName{}\textsubscript{DS-R1} & 791 (52.7\%) \da{10.0\%} & 998 (66.5\%) \da{13.2\%} & 1020 (68.0\%) \da{16.6\%} & 1021 (68.1\%) \da{19.2\%}\\
        & & \MethodName{}\textsubscript{unweighted} & 912 (60.8\%) \da{1.9\%} & 1159 (77.3\%) \da{2.4\%} & 1234 (82.3\%) \da{2.3\%} & 1306 (87.1\%) \da{0.2\%}\\
        \midrule
        \multirow{8}{*}{\DatasetName{}\textsubscript{CT}} & - & \textbf{\MethodName{}} & \textbf{183 (42.4\%)} & \textbf{259 (60.0\%)} & \textbf{285 (66.0\%)} & \textbf{292 (67.6\%)}\\
        \cmidrule(lr){2-7}
        & \multirow{2}{*}{Collection} & \MethodName{}\textsubscript{w/o stmt} & 151 (35.0\%) \da{7.4\%} & 211 (48.8\%) \da{11.2\%} & 220 (50.9\%) \da{15.1\%} & 224 (51.9\%) \da{15.7\%}\\
        & & \MethodName{}\textsubscript{w/o test} & 83 (19.2\%) \da{23.2\%} & 131 (30.3\%) \da{29.7\%} & 138 (31.9\%) \da{\textbf{34.1\%}} & 153 (35.4\%) \da{\textbf{32.2\%}}\\
        \cmidrule(lr){2-7}
        & Annotation & \MethodName{}\textsubscript{num} & 36 (8.3\%) \da{\textbf{34.1\%}} & 104 (24.1\%) \da{\textbf{35.9\%}} & 141 (32.6\%) \da{33.4\%} & 184 (42.6\%) \da{25.0\%}\\
        \cmidrule(lr){2-7}
        & \multirow{4}{*}{Voting} & \MethodName{}\textsubscript{Claude} & 145 (33.6\%) \da{8.8\%} & 154 (35.6\%) \da{24.4\%} & 154 (35.6\%) \da{30.4\%} & 154 (35.6\%) \da{32.0\%}\\
        & & \MethodName{}\textsubscript{DS-V3} & 147 (34.0\%) \da{8.4\%} & 166 (38.4\%) \da{21.6\%} & 167 (38.7\%) \da{27.3\%} & 167 (38.7\%) \da{28.9\%}\\
        & & \MethodName{}\textsubscript{DS-R1} & 168 (38.9\%) \da{3.5\%} & 180 (41.7\%) \da{18.3\%} & 181 (41.9\%) \da{24.1\%} & 181 (41.9\%) \da{25.7\%}\\
        & & \MethodName{}\textsubscript{unweighted} & 179 (41.4\%) \da{1.0\%} & 242 (56.0\%) \da{4.0\%} & 258 (59.7\%) \da{6.3\%} & 290 (67.1\%) \da{0.5\%}\\
        \bottomrule
    \end{tabular}
    \label{tab:RQ2}
    \vspace{-0.3cm}
\end{table*}

\subsection{RQ2: How effective are the design choices of \MethodName{}?}

To address this research question, we conduct an ablation study on \DatasetName{} by removing each component from \MethodName{} across its three stages, resulting in several variants.

\begin{itemize}[leftmargin=*]
    \item \textbf{Assignment-Specific Information Collection (\textit{Collection})}: \MethodName{} leverages problem statements (\textit{stmt}), test outcomes (\textit{test}), and reference programs (\textit{ref}) to enhance LLM's fault localization effectiveness. To assess the contribution of each type of information, we create three variants: \MethodName{}\textsubscript{w/o stmt}, \MethodName{}\textsubscript{w/o test}, and \MethodName{}\textsubscript{w/o ref}, each excluding one type of information.
    \item \textbf{Fault Location Annotation (\textit{Annotation})}: \MethodName{} prompts the LLM to annotate faulty lines instead of predicting line numbers. To examine the effectiveness of this strategy, we create a variant, \MethodName{}\textsubscript{num}, which predicts line numbers directly.
    \item \textbf{Weighted Multi-Model Voting (\textit{Voting})}: \MethodName{} determines the suspiciousness of each line by aggregating annotation results from Claude 3.7 Sonnet (\textit{Claude}), DeepSeek-V3 (\textit{DS-V3}), and DeepSeek-R1 (\textit{DS-R1}) via a weighted voting strategy. To examine the effectiveness of this strategy, we create three single-model variants (\MethodName{}\textsubscript{Claude}, \MethodName{}\textsubscript{DS-V3}, and \MethodName{}\textsubscript{DS-R1}) that replace the multi-model setup with each individual model, and one unweighted variant (\MethodName{}\textsubscript{unweighted}) that assigns a weight of $1$ to all annotated lines.
\end{itemize}

The results are shown in \autoref{tab:RQ2}, indicating that all components within \MethodName{} contribute positively to overall performance. Specifically:
\subsubsection{Assignment-Specific Information Collection} Excluding any type of information results in drops across all top-$k$ metrics on both datasets. Specifically, when excluding test outcomes, the top-1 accuracy drops from 62.7\% to 47.7\% on \DatasetName{}\textsubscript{DS} and from 42.4\% to 19.2\% on \DatasetName{}\textsubscript{CT}, representing the largest drop among all types of information. These results indicate that test outcomes are effective in improving fault localization, possibly by offering clues about the causes of test failures. This is especially valuable for programs in \DatasetName{}\textsubscript{CT}, which are longer and more complex than those in \DatasetName{}\textsubscript{DS}.
When excluding problem statements or reference programs, the top-$k$ metrics also drop, though less significantly, demonstrating that problem statements and reference programs are also effective, possibly by aiding in understanding the intended behavior of the program and guiding toward identifying the faults.

\subsubsection{Fault Location Annotation} Predicting line numbers directly results in significant drops across all top-$k$ metrics on both datasets. Specifically, the top-1 accuracy drops from 62.7\% to 15.8\% on \DatasetName{}\textsubscript{DS} and from 42.4\% to 8.3\% on \DatasetName{}\textsubscript{CT}, representing the largest drop among all components. These results reveal that the inherent limitations of LLMs in numerical understanding and processing severely hinder their fault localization performance, and our fault location annotation strategy effectively mitigates these limitations.

\subsubsection{Weighted Multi-Model Voting} Using individual models results in drops across all top-$k$ metrics on both datasets compared to the multi-model setup. Specifically, even when using the best-performing model (DeepSeek-R1), the top-1 accuracy drops from 62.7\% to 52.7\% on \DatasetName{}\textsubscript{DS} and from 42.4\% to 38.9\% on \DatasetName{}\textsubscript{CT}.
After analyzing the results, we observe that individual models consistently identify fewer faulty lines compared to the multi-model setup, resulting in more missed actual faulty lines. This is further evidenced by the slower growth rate of top-$k$ metrics as $k$ increases ($k\ge 3$) for individual models, particularly on the more challenging \DatasetName{}\textsubscript{CT} dataset. These findings indicate that while individual models may be powerful, each exhibits unique strengths and weaknesses. Our voting strategy effectively combines these complementary strengths, leading to superior performance.
Additionally, using the unweighted voting strategy also results in a performance drop. Specifically, the drops in top-1, top-3, and top-5 are more significant than the drop in top-10, indicating that our weighted voting strategy more effectively prioritizes actual faulty lines in higher-ranked positions. 

\vspace{1mm}
\begin{custommdframed}
\textbf{Answer to RQ2:} Removing any component from \MethodName{} results in drops across all top-$k$ metrics on both datasets, demonstrating each component's positive contribution to overall performance. Notably, \textit{Fault Location Annotation} proves the most effective component, highlighting its critical contribution to accurately identifying faulty lines.
\end{custommdframed}
\vspace{1mm}

\subsection{RQ3: How well does \MethodName{} generalize to general-purpose software codebases?}

To address this research question, we evaluate \MethodName{} on the widely used Defects4J~\cite{just2014defects4j} benchmark. Our evaluation follows the settings of the empirical study by \citet{wu2023large}, which was conducted within faulty functions. Since problem statements and reference programs are not available in Defects4J, we leverage only the test outcomes as auxiliary information. Specifically, we leverage the source code of failing unit tests and their corresponding error logs.

The results are shown in \autoref{tab:RQ3}. \MethodName{} outperforms all baseline methods across all top-$k$ metrics, with particularly notable improvements in top-1, top-3, and top-5, which are more critical for practical applications. In particular, \MethodName{} achieves a top-1 accuracy of 53.7\%, significantly higher than the best-performing baseline, ChatGPT-4 (Log). These results demonstrate that although \MethodName{} is primarily designed for programming assignments, it can effectively generalize to general-purpose software codebases beyond educational contexts.

\begin{table}[t]
    \centering
    \setlength{\abovecaptionskip}{0.1cm}
    \caption{Fault localization results on Defects4J.}
    \resizebox{\linewidth}{!}{\begin{tabular}{lcccc}
        \toprule
        \textbf{Method} & \textbf{Top-1} & \textbf{Top-3} & \textbf{Top-5} & \textbf{Top-10}\\
        \midrule
        FuseFL & 140 (34.3\%) & 201 (49.3\%) & 235 (57.6\%) & 283 (69.4\%)\\
        ChatGPT-4 (Log) & 171 (41.9\%) & 262 (64.2\%) & 285 (69.9\%) & 321 (78.7\%)\\
        Ochiai & 95 (23.3\%) & 190 (46.6\%) & 251 (61.5\%) & 294 (72.1\%)\\
        \textbf{\MethodName{}} & \textbf{219 (53.7\%)} & \textbf{297 (72.8\%)} & \textbf{318 (77.9\%)} & \textbf{324 (79.4\%)}\\
        \bottomrule
    \end{tabular}}
    \label{tab:RQ3}
    \vspace{-0.3cm}
\end{table}

\vspace{1mm}
\begin{custommdframed}
\textbf{Answer to RQ3:} \MethodName{} outperforms all baseline methods across all top-$k$ metrics on Defects4J, demonstrating its generalization capability to general-purpose software codebases beyond educational contexts.
\end{custommdframed}
\vspace{1mm}

\subsection{RQ4: How helpful are the results of \MethodName{} in assisting automated program repair?}
To address this research question, we conduct experiments on \DatasetName{}\textsubscript{DS} to investigate whether \MethodName{}'s fault localization results, \textit{i.e.}, identified faulty lines and their explanations, enhance the bug-fixing process when integrated with APR tools. Specifically, we implement an LLM-based APR pipeline using Claude 3.7 Sonnet~\cite{anthropic2025claude} as the repair engine, and conduct experiments under the following two settings:
\begin{itemize}[leftmargin=*]
    \item \textbf{Plain APR}: The LLM is provided with the faulty program, problem statement, and test outcomes, and is instructed to repair the program.
    \item \textbf{\MethodName{}-assisted APR}: In addition to the inputs in plain APR, the LLM is also provided with \MethodName{}'s results, including the identified faulty lines and their explanations.
\end{itemize}
We evaluate the effectiveness of APR using the following two metrics:
\begin{itemize}[leftmargin=*]
    \item \textbf{The number of fixed programs (\textit{\# Fixed})}: A program is considered \emph{fixed} if the repaired program \emph{passes all tests}.
    \item \textbf{The number of improved programs (\textit{\# Improved})}: A program is considered \emph{improved} if the repaired program \textit{passes more tests} than the original program, even if not passing all tests.
\end{itemize}

The results are shown in \autoref{tab:RQ4}. With the assistance of \MethodName{}'s results, the proportion of fixed programs increases from 57.2\% to 60.0\%, while the proportion of improved programs increases from 69.1\% to 71.8\%. These results demonstrate the helpfulness of \MethodName{}'s results in assisting APR.

To delve deeper into the role of \MethodName{}'s results in APR, we constructed a Sankey diagram (\autoref{fig:RQ4_Sankey}) illustrating the detailed repair results with and without the assistance of \MethodName{}'s results. The results reveal that a substantial number of failed fixes are either fixed or partially improved with the assistance of \MethodName{}'s results, further demonstrating its helpfulness. However, there are also a few programs that are fixed or partially improved failed to be fixed when \MethodName{}'s results are provided. We speculate that these failures may stem from the false positives in \MethodName{}'s results, which could misdirect the LLM into introducing erroneous edits to the correct code.

\begin{table}[t]
    \centering
    \setlength{\abovecaptionskip}{0.1cm}
    \caption{Results of \MethodName{}'s helpfulness in assisting APR on \DatasetName{}\textsubscript{DS}.}
    \begin{tabular}{lcc}
        \toprule
        \textbf{Setting} & \textbf{\# Fixed} & \textbf{\# Improved} \\
        \midrule
        Plain APR & 858 (57.2\%) & 1036 (69.1\%)\\
        \textbf{\MethodName{}-assisted APR} & \textbf{900 (60.0\%)} & \textbf{1077 (71.8\%)}\\
        \bottomrule
    \end{tabular}
    \label{tab:RQ4}
    \vspace{-0.3cm}
\end{table}

\begin{figure}[h]
    \centering
    \setlength{\abovecaptionskip}{0.1cm}
    \includegraphics[width=\linewidth]{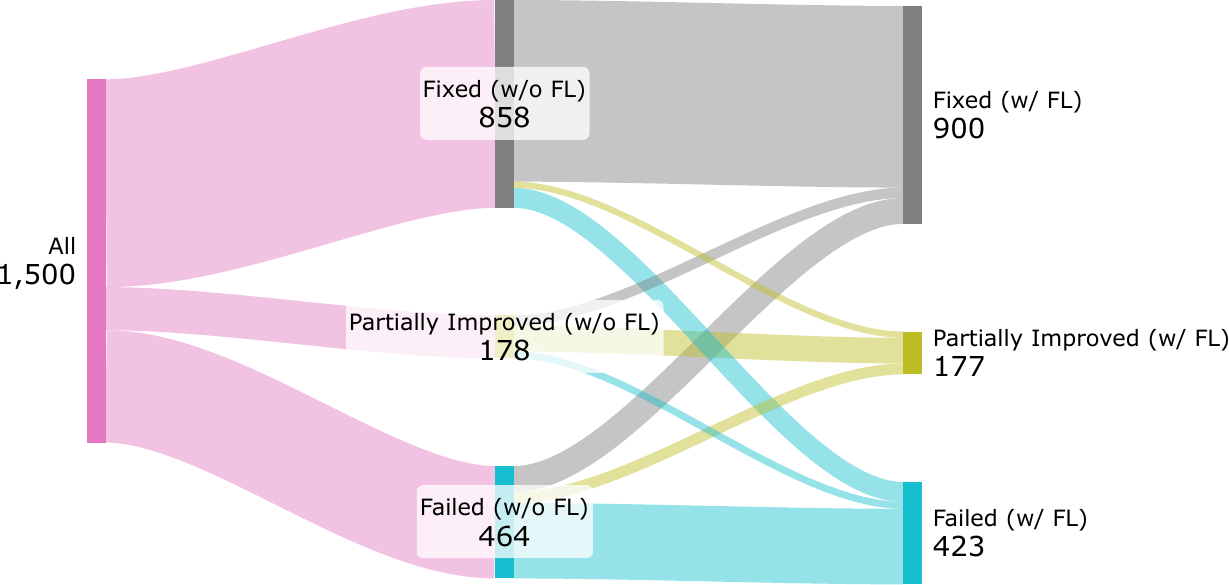}
    \caption{APR results with and without the assistance of \MethodName{}.}
    \label{fig:RQ4_Sankey}
    \vspace{-0.3cm}
\end{figure}

\vspace{1mm}
\begin{custommdframed}
\textbf{Answer to RQ4:} With the assistance of \MethodName{}'s fault localization results, both the number of fixed and improved programs increase, demonstrating that \MethodName{}'s results can provide valuable guidance for automated program repair.
\end{custommdframed}
\vspace{1mm}

\section{Discussion}

\subsection{Manual Analysis}

The correctness and informativeness of the explanations greatly affect their educational value. To assess these qualities, we manually analyzed the explanations generated by \MethodName{}. Specifically, we randomly sampled 65 cases from \DatasetName{}\textsubscript{DS} and 55 cases from \DatasetName{}\textsubscript{CT}, excluding the cases where \MethodName{} failed to localize the actual faulty lines within the top 5 most suspicious lines. The sample sizes were chosen to achieve a 90\% confidence level with a 10\% margin of error. For each sampled case, we manually checked the explanation's \textbf{correctness} (categorized as \textit{true} or \textit{false}) and rated its \textbf{informativeness} using a 7-point Likert scale~\cite{likert1932technique} (1 = \textit{minimally informative}, 7 = \textit{highly informative}), with higher scores indicating greater informativeness. The analysis was carried out by one of the authors, who possesses over 4 years of C++ and Java programming experience.

\begin{figure}[t]
    \centering
    \includegraphics[scale=0.7]{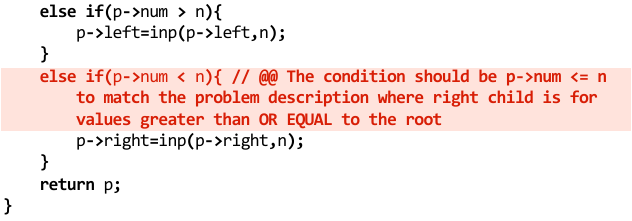}
    \caption{An example of a highly informative explanation (score = 7).}
    \label{fig:Informativeness1}
    \vspace{-0.3cm}
\end{figure}

\begin{figure}[h]
    \centering
    \includegraphics[scale=0.7]{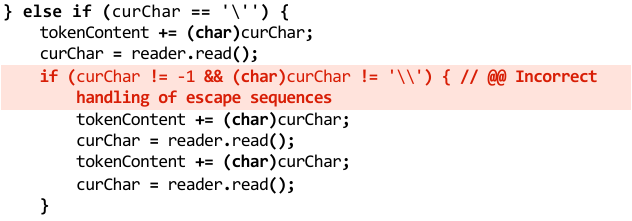}
    \caption{An example of a less informative explanation  (score = 3).}
    \label{fig:Informativeness2}
    \vspace{-0.3cm}
\end{figure}

Our analysis showed that \MethodName{} generated correct explanations in 59 out of 65 \DatasetName{}\textsubscript{DS} cases (90.8\%) and in 52 out of 55 \DatasetName{}\textsubscript{CT} cases (94.5\%). The average informativeness scores were 6.47 and 6.29, respectively. These results demonstrate \MethodName{}'s ability to generate explanations that are both correct and highly informative, highlighting its educational value.
\autoref{fig:Informativeness1} shows an example of a highly informative explanation (score = 7). In this case, the explanation clearly indicates that \textit{the \texttt{<} operator in the \texttt{if} condition should be changed to \texttt{<=} to align with the problem description: values greater than \textbf{or equal to} the current node should be in the right subtree}. This explanation is actionable for students to understand and fix the error. In contrast, \autoref{fig:Informativeness2} exemplifies a less informative explanation (score = 3). Here, the explanation vaguely notes that \textit{the handling of escape sequences is incorrect}, failing to specify the nature of the error or suggest a fix, limiting its usefulness for students.

\subsection{Case Study of Failures}

Despite the effectiveness of \MethodName{}, it may still fail to accurately localize the actual faulty lines within the top 5 most suspicious lines in certain scenarios. To better understand these limitations, we conducted case studies on \DatasetName{} and identified the following recurring failure patterns:

\subsubsection{Correct Diagnosis but Incorrect Localization} 
\begin{figure}[t]
    \centering
    \includegraphics[scale=0.7]{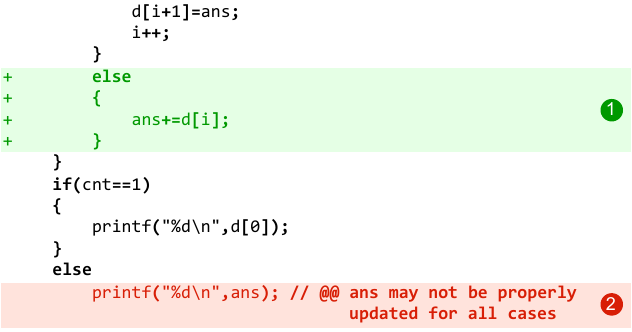}
    \caption{An example of correct diagnosis but incorrect localization.}
    \label{fig:Failure1}
    \vspace{-0.3cm}
\end{figure}

In several cases, \MethodName{} correctly diagnosed the root cause of the error but incorrectly identified the location. As illustrated in \autoref{fig:Failure1}, the error arises from a missing update to the \texttt{ans} variable (\ding{182}). While \MethodName{} correctly diagnosed the root cause of the error in its explanation, it incorrectly annotated the line where \texttt{ans} is printed (\ding{183}). We believe that while these cases are classified as failures under the current strict ground-truth criteria (which prioritize precise localization), they can still guide students toward understanding the errors in their programs. 

\subsubsection{Failure on Complex Programs} 
\begin{figure}[h]
    \centering
    \includegraphics[scale=0.7]{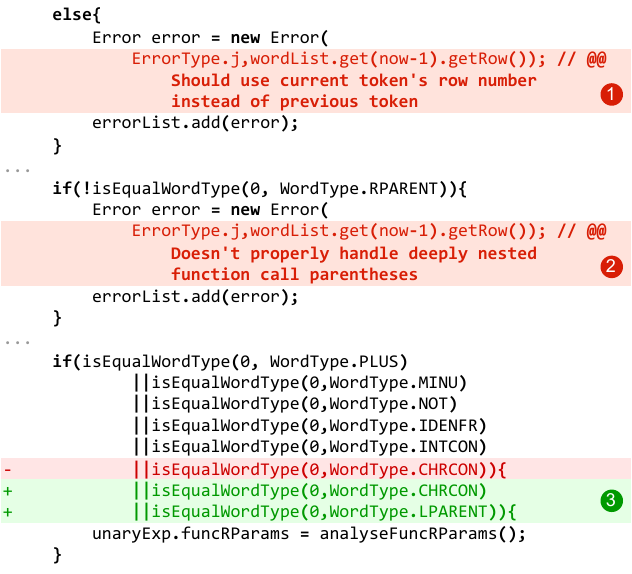}
    \caption{An example of failure on a complex program.}
    \label{fig:Failure2}
    \vspace{-0.3cm}
\end{figure}

In some scenarios, \MethodName{} exhibited total failure in identifying the error. As shown in \autoref{fig:Failure2}, the program does not properly handle parenthesized expressions when parsing function arguments (\ding{184}). \MethodName{} was unable to navigate the complexities of the program and failed to identify the subtle error buried within its 1,900 lines of code. As a result, it reported several incorrect faulty lines with misleading explanations (\ding{182},\ding{183}).
These cases highlight limitations in \MethodName{}'s ability to reason about semantics in large-scale programs, underscoring both the challenge and critical need for fault localization techniques to advance robustness against real-world software complexity.

\subsubsection{Failure in Suspicious Line Ranking}
Additionally, there are cases where \MethodName{} correctly identified ground-truth faulty lines but ranked them outside the top-5 most suspicious lines, resulting in their classification as failures.
These cases underscore the need to enhance LLM's semantic understanding and reasoning capabilities, as well as to refine the voting and ranking strategies.

\subsection{Threats to Validity}

\textbf{Threats to internal validity} relate to the reproducibility of \MethodName{}'s results, given the stochastic nature of sampling from LLMs. Although we mitigate this issue by performing multiple rounds of fault location annotation and aggregating the results via a weighted voting strategy, completely eliminating uncertainty remains challenging. As a result, the reproducibility of \MethodName{}'s results may still be affected to some extent.

\textbf{Threats to external validity} relate to the availability of auxiliary information leveraged by \MethodName{}.
\MethodName{} leverages auxiliary information such as problem statements, test outcomes, and reference programs retrieved from historical submissions. However, the availability and quality of such information cannot always be guaranteed. In scenarios where the auxiliary information is limited or unreliable, \MethodName{}'s effectiveness may be adversely affected.

\section{Conclusion and Future Work}

In this paper, we propose \MethodName{}, a fine-grained, explainable fault localization method tailored for programming assignments. \MethodName{} leverages rich contextual information specific to programming assignments to guide LLMs in identifying faulty code lines. By prompting LLMs to annotate faulty code lines with detailed natural language explanations, our method enhances both localization accuracy and educational value. To further improve reliability, we introduce a weighted multi-model voting strategy that aggregates the results of multiple LLMs. Extensive experimental results show that \MethodName{} outperforms state-of-the-art fault localization baselines on both our newly constructed programming assignment benchmark (\DatasetName{}) and the widely used Defects4J benchmark for general-purpose codebases. 

In the future, we plan to enhance \MethodName{}'s semantic understanding and reasoning capabilities, particularly for large-scale programs. We will also refine the voting and ranking strategies by better distinguishing between different types of errors. Finally, we plan to investigate our method's generalization capability across diverse programming languages and educational contexts with varying assignment types.

\section*{Acknowledgement}
This research is supported by the National Natural Science Foundation of China Grants Nos. 62302021, 62177003, 62332001, the Fundamental Research Funds for the Central Universities (Grant No. JK2024-28),
and Beijing Advanced Innovation Center for Future Blockchain and Privacy Computing.

\bibliography{references}

\end{document}